\documentclass{elsartL}
\usepackage{graphicx}
\usepackage{amssymb}
\usepackage{amsmath}
\usepackage{mathrsfs}
\usepackage{physics}
\usepackage{mathtools}% http://ctan.org/pkg/mathtools
\newcommand{\fsl}[1]{\ensuremath{\mathrlap{\!\not{\phantom{#1}}}#1}}% \fsl{<symbol>}
\newcommand{\avg}[1]{\left< #1 \right>} % For a nice display of the average value
\begin{document}

\begin{frontmatter}
\title{The contribution of the $\rho^0 - \omega$ interference to the violation of the isospin invariance in the $\pi N$ system}
\author{E.~Matsinos{$^*$}}

\begin{abstract}
It is generally believed that the $\rho^0 - \omega$ interference is the dominant mechanism underlying the violation of the isospin invariance in the $N N$ system. Assessed in this paper is the role which the mechanism might 
play in a potential departure from the isospin symmetry in the two $\pi^\pm p$ elastic-scattering reactions.
\end{abstract}
%13.75.Gx: neutron-pion interactions, nucleon-meson interactions, nucleon-pion interactions, pion-baryon reactions, proton-pion interactions
%25.80.Dj: elastic scattering (meson-induced reactions), elastic scattering (pion-nucleus)
%25.80.Gn: charge-exchange reactions (pion), pion absorption and capture
%11.30.-j: symmetry in theory of fields and particles, conservation laws fields and particles
{$^*$}{E-mail: evangelos (dot) matsinos (at) sunrise (dot) ch}
\end{frontmatter}

\section{\label{sec:Introduction}Introduction}

In Hadronic Physics, the isotopic (isobaric) spin or isospin ($I$) is a quantum number, on the basis of which the hadrons may be categorised into families (multiplets) containing particles with (ideally) identical properties 
as far as the strong interaction is concerned. The $2I+1$ members of each multiplet are thought of as representing different states of one (notional) particle; for example, the proton ($p$) and the neutron ($n$) are regarded 
as two manifestations of one fictitious particle, of the `nucleon' \cite{heisenberg1932}. The nucleon is assigned a total isospin of $I=1/2$, the third projection of which ($I_3$) specifies its type/manifestation, as a 
proton when $I_3=+1/2$ or as a neutron when $I_3=-1/2$. The isospin invariance (of the strong interaction) had originally been thought of as exact, in which case the members of the isospin multiplets would be treated (by 
the strong interaction) on equal footing.

Applied to pions and nucleons, of central interest in this paper, the isospin invariance ensures that the (strong) forces between the three pion types ($\pi^+$, $\pi^0$, and $\pi^-$) and the two nucleon types ($p$ and $n$) 
are the same, regardless of the interacting particles. In a hypothetical world where the electromagnetic (EM) interaction could be switched off and the isospin invariance would be exact, the scattering amplitudes of the various 
$\pi N$ reactions would be determined by merely following the standard formalism of the isospin structure (ladder operators, orthogonality conditions).

The violation of the isospin invariance in the nucleon-nucleon ($N N$) interaction was established shortly before the introduction of the current theory of the strong interaction, i.e., of Quantum Chromodynamics (QCD). 
Referring to low energy, the strong part of the $N N$ interaction is characterised by three scattering lengths, corresponding to the $^1S_0$ states $p p$, $n n$, and $n p$. If the charge independence (this term is used in 
the $N N$ domain as a synonym for isospin invariance) would hold, the values of these three quantities would be equal. In reality, after the removal of the EM effects, their values are \cite{miller2006}:
\begin{equation}
a_{pp}=-17.3(4) \,\, {\rm fm}, \, \, \, a_{nn}=-18.8(3) \,\, {\rm fm}, \, \, \, a_{np}=-23.77(9) \,\, {\rm fm} \, \, \, .
\end{equation}
(In Ref.~\cite{miller2006}, these scattering lengths carry the superscript `N', indicating that they are nuclear quantities, i.e., obtained after the removal of the EM effects.) Obviously, these values violate the charge 
independence by about $27 \%$ and the charge symmetry by about $8 \%$. The differences
\begin{equation}
\Delta a_{CD}=(a_{pp}+a_{nn})/2-a_{np}=5.7(3) {\rm fm}
\end{equation}
and
\begin{equation}
\Delta a_{CSD}=a_{pp}-a_{nn}=1.5(5) {\rm fm}
\end{equation}
translate the significance of the two effects into number of $\sigma$'s in the normal distribution.

The explanation of the (large) isospin-breaking effects in the $N N$ system served as the main motivation for the study of the $\rho^0 - \omega$ interference phenomenon during the last four decades of the twentieth century. 
Hypothesised by Glashow in 1961 \cite{glashow1961}, the interference between these two states was first investigated in a paper by Coleman and Glashow in 1964 \cite{coleman1964}. Within a few years of that study, the first 
experimental evidence of the phenomenon emerged (see works [1-4] in Ref.~\cite{allcock1970}). By the late 1970s, it was known that the violation of the isospin invariance in the $N N$ system (also referred to as `nuclear 
charge asymmetry') could mostly be accounted for by the $\rho^0 - \omega$ interference phenomenon (e.g., see works [1,2] in Ref.~\cite{coon1987}). Since then, the interference between the two states has been investigated 
in a multitude of studies, e.g., see Refs.~\cite{goldman1992,piekarewicz1993,krein1993,cohen1995,oconnell1997,coon1998,wang2000,yan2002,azimov2003,azimov2003_2} (this list is not exhaustive).

Given that the $\rho$ meson is an $I^G (J^{PC}) = 1^+ (1^{--})$ object, whereas the $\omega$ meson is identified as an $I^G (J^{PC}) = 0^- (1^{--})$ state, the coupling between these two particles implies a $\Delta I=1$ 
transition vertex, i.e., it explicitly violates the isospin invariance. The coupling of the $\omega$ meson to two pions is forbidden due to the G-parity violation. Nevertheless, inspection of the $\omega$-meson decay 
modes (e.g., see the present-day data by the Particle-Data Group (PDG) \cite{pdg2016}) reveals that the branching ratio of the $\omega$-meson decay into the $\pi^+ \pi^-$ channel, albeit suppressed, is significantly 
non-zero: it amounts to about $1.5 \%$.

The procedure for the inclusion of the effects of the $\rho^0 - \omega$ interference in scattering processes should not be thought of as clear-cut and `beyond doubt'. In many works dealing with $\omega$-meson-related 
effects, the direct coupling of the $\omega$ meson to the pion is ignored. For instance, the $\rho^0 - \omega$ mixing amplitude was directly associated in Ref.~\cite{coon1987} with the branching ratio of the $\omega$ 
meson into a pair of pions. Such an approach assumes that the decay mode $\omega \to \pi^+ \pi^-$ is entirely due to the $\rho^0 - \omega$ interference phenomenon and that the G-parity holds. On the other hand, Maltman, 
O'Connell, and Williams \cite{maltman1996} maintained that both
\begin{itemize}
\item the $\rho^0 - \omega$ interference mechanism (and the consequent violation of the isospin invariance) and 
\item the direct coupling of the $\omega$ meson to two pions (and the consequent violation of the G-parity) 
\end{itemize}
must be taken into account when addressing the importance of the $\omega$-meson-related effects. Nevertheless, unable to set forth a reliable procedure to disentangle the contributions from the two aforementioned mechanisms 
to the observed branching ratio $\omega \to \pi^+ \pi^-$, the direct coupling of the $\omega$ meson to the pion will be ignored also in this work, and the non-zero branching ratio $\omega \to \pi^+ \pi^-$ will entirely 
be attributed to the $\rho^0 - \omega$ interference.

The violation of the isospin invariance in the interactions between pions and nucleons has been reported in a number of studies since the mid 1990s, see Refs.~\cite{gibbs1995,matsinos1997,matsinos2006,matsinos2013}. In a 
recent paper \cite{matsinos2017}, it was shown that the isospin-breaking effects find their way to the output of dispersion-relation analyses assuming \emph{ab initio} the fulfilment of the isospin invariance, as the 
case is with the current solution (WI08) of the SAID group \cite{arndt2006}.

Despite the fact that overwhelming evidence had been produced that the isospin invariance is broken in the $N N$ system, the first reports \cite{gibbs1995,matsinos1997} of similar effects in the $\pi N$ system were 
received with scepticism\footnote{This attitude might have been moulded by studies of the phenomenon conducted within the framework of the Chiral-Perturbation Theory; the general conclusion of such studies is that 
the magnitude of the corresponding effects in the $\pi N$ interaction should not exceed the percent level, e.g., see Ref.~\cite{hoferichter2010} and the references therein.}. However, the $\pi N$ interaction is a basic 
element in the modelling of the $N N$ system in the framework of meson-exchange models of the strong interaction. For example, the CD-Bonn potential \cite{machleidt2001} rests upon $t$-channel Feynman graphs (hereafter, 
simply `graphs') with $\pi$, $\eta$, $\rho$, $\omega$, and (two) $\sigma$ mesons as intermediate states. In such a scheme, the possibility arises that part of the isospin-breaking effects, observed in the $N N$ interaction, 
could cascade down to the $\pi N$ system.

Although the theoretical frameworks, upon which Refs.~\cite{gibbs1995,matsinos1997} were based, were different (the first used effective potentials, the second a hadronic model), they followed similar strategies. Reference 
\cite{gibbs1995}
\begin{itemize}
\item extracted (from low-energy data) the scattering amplitudes of the three low-energy reactions, i.e., of the two elastic-scattering (ES) processes and of the $\pi^- p$ charge-exchange (CX) reaction 
$\pi^- p \to \pi^0 n$,
\item employed the triangle identity to derive (from the scattering amplitudes of the two ES processes) a (notional) scattering amplitude for the $\pi^- p$ CX reaction (to be called `reconstructed'), and 
\item compared the extracted and reconstructed scattering amplitudes of the $\pi^- p$ CX reaction. The amount of the dissimilarity in this comparison is a measure of the violation of the triangle identity, hence of 
the isospin invariance in the $\pi N$ system.
\end{itemize}
On the other hand, Ref.~\cite{matsinos1997} 
\begin{itemize}
\item obtained estimates for the parameters of an isospin-invariant hadronic model from a fit to (compared to Ref.~\cite{gibbs1995}, a slightly larger set of) low-energy (pion laboratory kinetic energy $T \leq 100$ MeV) 
$\pi^\pm p$ ES data (differential cross sections, analysing powers, and partial-total/total cross sections) and 
\item investigated the reproduction of the low-energy $\pi^- p$ CX data on the basis of these parameter values.
\end{itemize}
The triangle identity is incorporated in the approach of Ref.~\cite{matsinos1997} at the level of the derivation of the observables for the three low-energy reactions from the hadronic phase shifts obtained from the model 
parameters.

These two independent analyses arrived at similar conclusions: provided that the bulk of the low-energy $\pi N$ data is reliable and that any missing EM effects are small, the isospin invariance is violated in the low-energy 
$\pi N$ interaction, at a level of about $5$ to $10 \%$ (effect in the scattering amplitude): most of the effect is to be attributed to the $s$-wave part of the scattering amplitude. (Subsequent work demonstrated that the effect 
at low energy is mostly traced to the hadronic phase shift $S_{11}$ \cite{matsinos2013}.)

\begin{figure}
\begin{center}
\includegraphics [width=15.5cm] {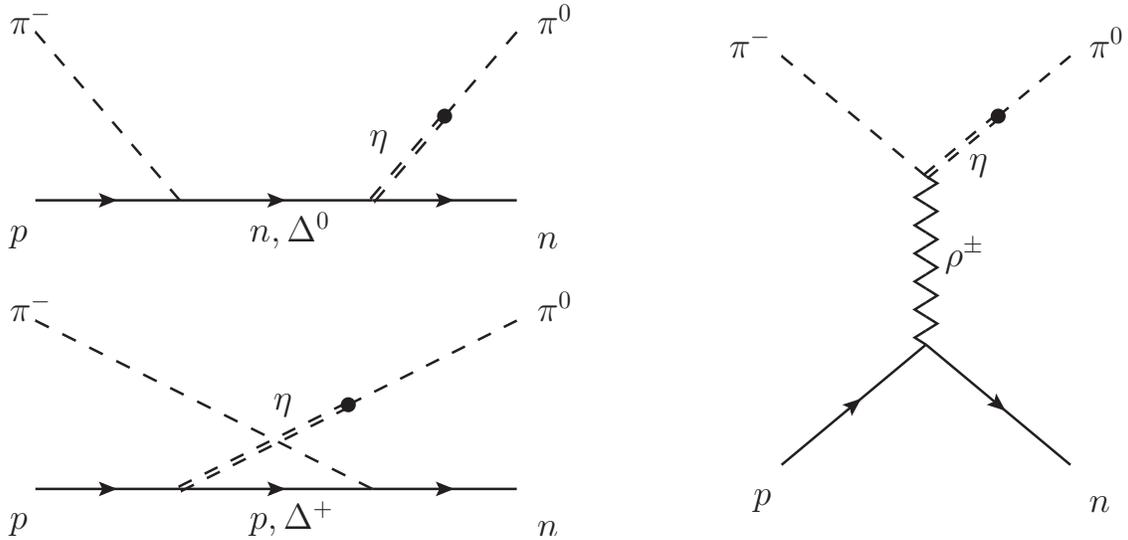}
\caption{\label{fig:IsospinBreakingEtaPi0}Feynman graphs of the ETH model involving the $\pi^0 - \eta$ interference, a potential mechanism for the violation of the isospin invariance in the hadronic part of the $\pi N$ 
interaction in the $\pi^- p$ charge-exchange reaction.}
\vspace{0.35cm}
\end{center}
\end{figure}

It is generally believed that the isospin-breaking effects (relevant to systems of pions and nucleons) in QCD arise from the difference of the masses of the up ($u$) and down ($d$) quarks. In the modelling of the strong 
interaction via meson-exchange graphs, there are three ways in which isospin-breaking effects might enter the hadronic part of the scattering amplitude:
\begin{itemize}
\item via differences in the coupling constants and vertex factors,
\item via mass differences between the members of the isospin multiplets, and
\item via the introduction of isospin-breaking graphs.
\end{itemize}
Regarding the last option, Cutkosky proposed in the late 1970s the $\pi^0 - \eta$ interference as a potential mechanism for the departure from the triangle identity in the $\pi N$ system \cite{cutkosky1979}, see 
Fig.~\ref{fig:IsospinBreakingEtaPi0}.

\begin{figure}
\begin{center}
\includegraphics [width=7.75cm] {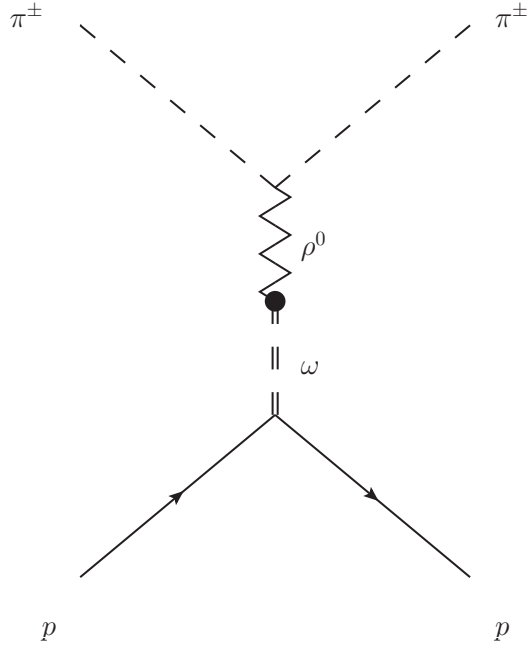}
\caption{\label{fig:IsospinBreakingRhoOmega}Feynman graph of the ETH model involving the $\rho^0 - \omega$ interference, a potential mechanism for the violation of the isospin invariance in the hadronic part of the $\pi N$ 
interaction in the elastic-scattering reactions $\pi^\pm p \to \pi^\pm p$.}
\vspace{0.35cm}
\end{center}
\end{figure}

The approach, followed in Refs.~\cite{gibbs1995,matsinos1997}, might be taken to suggest that the \emph{$\pi^- p$ CX reaction} is to blame for the violation of the isospin invariance. The aim in this paper is to investigate 
whether the $\rho^0 - \omega$ interference could lead to sizeable isospin-breaking effects in the \emph{two ES reactions} (see Fig.~\ref{fig:IsospinBreakingRhoOmega}). Significant contributions to the $\pi N$ scattering 
amplitude could provide an explanation for a puzzling discrepancy in the analysis of the low-energy $\pi N$ data in this research programme, namely for the mismatch between a) the $\pi^- p$ ES length $a^{cc}$ obtained via 
the extrapolation to the $\pi N$ threshold (zero kinetic energy of the incident pion) of the $\pi^- p$ ES amplitude (determined from experimental data above threshold) and b) the $a^{cc}$ value obtained directly at threshold, 
in the two pionic-hydrogen experiments conducted at the Paul Scherrer Institut (Villigen, Switzerland) between about 1995 and 2005.

\section{\label{sec:Rho}The coupling of the $\rho$ meson to the $\pi N$ system}

In effective (hadronic) models of the $\pi N$ interaction, the $t$-channel $I=J=1$ contribution is modelled via the $\rho$-meson-exchange graph. In the $\pi N$ model developed at the ETH (Zurich, Switzerland) during the 
1990s \cite{goudsmit1994,matsinos2014,ZRH17}, the $\pi \pi \rho$ Lagrangian density reads as:
\begin{equation} \label{eq:EQ01}
\Delta \mathscr{L}_{\pi \rho} = g_{\pi \pi \rho} \vec{\rho} \, ^\mu \cdot (\partial_\mu \vec{\pi} \times \vec{\pi} ) \, \, \, ,
\end{equation}
where $\vec{\pi}$ and $\vec{\rho} \, ^\mu$ denote the quantum fields of the pion and of the exchanged $\rho$ meson, respectively. On the other hand, the $\rho N N$ Lagrangian density reads as:
\begin{equation} \label{eq:EQ02}
\Delta \mathscr{L}_{\rho N} = - g_{\rho N N} \bar{\psi} \frac{\vec{\tau}}{2} \cdot \left( \gamma^\mu \vec{\rho} \, _\mu + \frac{\kappa_\rho}{2 m_p} \sigma^{\mu \nu} \partial_\mu \vec{\rho}_\nu \right) \psi \, \, \, ,
\end{equation}
where $\psi$ stands for the field of the nucleon, its isospin operator being represented by $\vec{\tau}/2$, and $\kappa_\rho$ determines the relative strength of the tensor coupling in the $\rho N N$ vertex. It is conventional 
to use the proton mass $m_p$ in the tensor coupling. The antisymmetric tensor $\sigma^{\mu \nu}$ is defined in terms of the Dirac matrices via the relation: $\sigma^{\mu \nu} = i [\gamma^\mu , \gamma^\nu] / 2$. In fact, the 
contributions of the $\rho$-exchange graph to the $\pi N$ scattering amplitude at low $4$-momentum transfer depend on one effective (Fermi-like) parameter - $G_\rho$ - obeying $G_\rho m_\rho^2 = g_{\pi \pi \rho} g_{\rho N N}$. 
The $\pi N$ scattering length $a^{cc}$ is nearly accounted for by the contribution of the $\rho$-meson-exchange graph to the invariant amplitude $B$ (see Section 2.3 of Ref.~\cite{matsinos2014} for the definitions of the 
invariant amplitudes $A$ and $B$); this contribution is proportional to $G_\rho$.

In his book \cite{hoehler1983}, H\"ohler cites a relation between the coupling $g_{\pi \pi \rho}$ and the decay width $\Gamma(\rho \to \pi^+ \pi^-)$ of the $\rho$ meson (relation A.8.29, p.~565), see also Eq.~(5) in 
Ref.~\cite{allcock1970}:
\begin{equation} \label{eq:EQ03}
\Gamma(\rho \to \pi^+ \pi^-) = \frac{g_{\pi \pi \rho}^2}{4 \pi} \frac{2}{3} \frac{q_\rho^3}{m_\rho^2} \, \, \, ,
\end{equation}
where $q_\rho$ denotes the magnitude of the $3$-momentum of the emitted pions in the centre-of-mass (CM) frame, at the energy equivalent to the $\rho$-meson mass $m_\rho$:
\begin{equation} \label{eq:EQ04}
q_\rho = \sqrt{\frac{m_\rho^2}{4}-m_c^2} \, \, \, ,
\end{equation}
and $m_c$ stands for the mass of the charged pion. Using the present-day values of the physical constants \cite{pdg2016}, one obtains from Eq.~(\ref{eq:EQ03}):
\begin{equation} \label{eq:EQ04_1}
g_{\pi \pi \rho} = 5.950 \pm 0.018 \, \, \, .
\end{equation}

In the remaining part of this section, issues relevant to the $\rho N N$ vertex will be addressed. For the coupling constant $g_{\rho N N}$, H\"ohler recommends the use of the dispersion-relation result \cite{hoehler1975}:
\begin{equation} \label{eq:EQ05}
\frac{g_{\rho N N}^2}{4\pi} \approx 2.2 \, \, \, .
\end{equation}
H\"ohler assessed the relative uncertainty of $g_{\rho N N}^2 / (4\pi)$ to $15 \%$ (see Chapter 2.4.2.2 of Ref.~\cite{hoehler1983}, p.~228). One thus obtains $g_{\rho N N} = 5.26 \pm 0.39$, a value which should be 
taken as corresponding to the $\rho$-meson pole; after comparing the strength of the $\rho$-meson couplings to the pion and to the nucleon (vector coupling), one observes no dramatic departure from universality (equal 
couplings). One may next obtain an estimate for the parameter $G_\rho$ (of the ETH model) from H\"ohler's values for the two coupling constants (assuming no $t$ dependence of the couplings\footnote{$t$ is the standard 
Mandelstam variable, denoting the square of the $4$-momentum transfer.}): $G_\rho = 52.1 \pm 3.9$ GeV$^{-2}$. The $G_\rho$ result from an updated analysis of the low-energy $\pi^\pm p$ ES data \cite{ZRH17} is: 
$G_\rho = 55.37 \pm 0.61$ GeV$^{-2}$; this value should be thought of as corresponding to $t=0$. The comparison of these two values reveals that the $t$ dependence of $G_\rho$ must be weak. This is not a new result; the 
same conclusion was drawn in all past phase-shift analyses (PSAs) of the low-energy $\pi N$ data with the ETH model.

One may advance one step further and obtain an estimate for $g_{\rho N N}$ from the values of $G_\rho$, $g_{\pi \pi \rho}$, and $m_\rho$. In the framework of the ETH model, one thus obtains from the fits to the low-energy 
$\pi^\pm p$ ES data:
\begin{equation} \label{eq:EQ05_1}
g_{\rho N N} = 5.593 \pm 0.064 \, \, \, .
\end{equation}
To conclude, the overall picture regarding the vector coupling of the $\rho$ meson to the nucleon appears satisfactory; the sign differences in $\Delta \mathscr{L}_{\pi \rho}$ and $\Delta \mathscr{L}_{\rho N}$ between 
Refs.~\cite{goudsmit1994,matsinos2014} and \cite{hoehler1983} are not important.

Concerning the tensor coupling in the $\rho N N$ vertex, the situation is not equally comforting. To start with, the ratio of the tensor-to-vector couplings (denoted as $\kappa_\rho$ in the ETH model) had been estimated 
by means of dispersion relations to about $6.6$ \cite{hoehler1975} (subsequent results, obtained from the $N N$ data, were not incompatible with this estimate). On the contrary, the $\kappa_\rho$ results, obtained with 
the ETH model from fits to the low-energy $\pi N$ data, have always been considerably smaller; the current result from the analysis of the ES data is $\kappa_\rho = 0.73 \pm 0.37$, whereas a slightly larger result has 
been obtained from the combined analysis of the $\pi^+ p$ and $\pi^- p$ CX data: $\kappa_\rho = 1.47 \pm 0.17$ \cite{ZRH17}. In short, the difference in the $\kappa_\rho$ values, between the dispersion-relation results 
and those obtained from the low-energy $\pi N$ data with the ETH model, appears unbridgeable.

At this point, one should mention that the $\kappa_\rho$ value, extracted from dispersion-relation analyses, applies to the $\rho$-meson pole; as the PSAs using the ETH model involve low-energy data, one could argue that 
it is more apt to compare the $\kappa_\rho$ estimates, obtained with the ETH model, with the vector-meson dominance (VMD) prediction (i.e., $\kappa_\rho \approx 3.71$ \cite{sakurai1969}), see also the discussion in Section 
III of Ref.~\cite{coon1987}. However, even in that case, the discrepancy is sizeable. Further comments on this issue will be made in the subsequent section.

Some remarks on the sign of the tensor coupling in Ref.~\cite{hoehler1983} are due; it is opposite to that of Eq.~(\ref{eq:EQ02}). When (in early 1993) the tensor coupling was included in the ETH model for the first time 
(with a sign opposite to H\"ohler's, i.e., as given in Eq.~(\ref{eq:EQ02})), a surprising result was obtained: the fitted $\kappa_\rho$ value came out positive. This result instigated a thorough examination of the original 
calculations and the comparison of the contributions (in particular, those involving the $\rho$-meson-exchange graph) to the standard invariant amplitudes $A$ and $B$. The conclusion of that investigation was that, in spite 
of the difference in the tensor coupling of the Lagrangian density $\Delta \mathscr{L}_{\rho N}$, the contributions of the $\rho$-meson-exchange graph to the $\pi N$ scattering amplitude, obtained in the two schemes (i.e., 
in the context of the ETH model and from H\"ohler's book), were identical. Although it is not straightforward to point out which of the definitions in the two schemes differ(s), it nevertheless emerged that the discrepancy 
in the two forms of the Lagrangian density $\Delta \mathscr{L}_{\rho N}$ was due to a convention\footnote{Of course, the alternative is that the sign in H\"ohler's book is simply a mistype.}; as the contributions of the 
$\rho$-meson-exchange graph to the partial-wave amplitudes, hence to the observables, are identical, any such differences are inessential.

As time went by, it was discovered that various forms of $\Delta \mathscr{L}_{\rho N}$ have been used in the literature. To mention a few differences, Refs.~\cite{coon1987,pearce1991} make use\footnote{There is a mistype 
in Eq.~(6a) of Ref.~\cite{coon1987}: the plus sign after $\gamma_\mu$ on the right-hand side (rhs) should not appear as a superscript.} of Eq.~(\ref{eq:EQ02}), whereas 
Refs.~\cite{duttaroy1969,lee1991,piekarewicz1993,krein1993,brown1994,machleidt2001} retained the sign of Eq.~(\ref{eq:EQ02}), but did not include the factor of $2$ in the definition of the isospin operator of the nucleon. 
Finally, Ref.~\cite{downum2006} used H\"ohler's sign, also omitting the factor of $2$. One cannot easily come up with arguments in favour of this plurality. Fortunately however, given that the $\rho$-meson contribution is 
the dominant piece in the isovector $s$-wave part of the $\pi N$ interaction, any essential differences would find their way to the important results, namely to the contributions to the observables, and would thus be readily 
noticeable. In the calculations involving the ETH model, the Lagrangian densities of Eqs.~(\ref{eq:EQ01},\ref{eq:EQ02}) have always been used. In the following, the appropriate factors (conforming to the conventions 
detailed in Ref.~\cite{matsinos2014}) will be applied to all quantities referred to or imported from other works; unless stated otherwise, all such values pertain to $t=0$, i.e., they are suitable for low-energy applications.

\section{\label{sec:Omega}The coupling of the $\omega$ meson to the nucleon}

The $\omega N N$ Lagrangian density reads as:
\begin{equation} \label{eq:EQ06}
\Delta \mathscr{L}_{\omega N} = - g_{\omega N N} \bar{\psi} \frac{1}{2} \left( \gamma^\mu \omega_\mu + \frac{\kappa_\omega}{2 m_p} \sigma^{\mu \nu} \partial_\mu \omega_\nu \right) \psi \, \, \, .
\end{equation}
Apart from a sign convention, this is the form used by Coon and Barrett \cite{coon1987}. To facilitate the comparison of the $g_{\rho N N}$ and $g_{\omega N N}$ values with works which omit the factor of $2$ from 
the definition of the isospin operator of the nucleon in Eq.~(\ref{eq:EQ02}), the factor of $1/2$ will be retained on the rhs of Eq.~(\ref{eq:EQ06}).

Three standard analyses of the $N N$ data have reported on the coupling constant $g_{\omega N N}$ and on the ratio $\kappa_\omega$ of Eq.~(\ref{eq:EQ06}); arranged in chronological order, these values are given in Table 
\ref{tab:Constants}. Due to the use of form factors in these analyses, some attention is needed in order to interpret correctly the reported results; one might miss this important point, in particular if one has not come 
across Ref.~\cite{brown1994} (p.~1732). The Nijmegen and the Paris $g_{\rho N N}$ and $g_{\omega N N}$ values correspond to $t=0$; in both analyses, form factors are applied to the meson-nucleon vertices at $t \neq 0$. 
For example, to obtain the Nijmegen estimates for the coupling constants $g_{\rho N N}$ and $g_{\omega N N}$ at the corresponding meson pole, one would need to multiply the quoted $g_\alpha^2/(4 \pi)$ values by $e^{t_\alpha / \Lambda^2}$, 
where $t_\alpha=m_\alpha^2$ and $\Lambda = 964.52$ MeV. On the contrary, the Bonn $g_{\rho N N}$ and $g_{\omega N N}$ values refer to the corresponding meson poles; as indicated on p.~24 of Ref.~\cite{machleidt2001}: 
``At each meson-nucleon vertex, a form factor is applied which has the analytical form:
\begin{equation} \label{eq:EQ06_0}
\mathscr{F}_\alpha (t) = \frac{\Lambda_\alpha^2 - m_\alpha^2}{\Lambda_\alpha^2 - t}
\end{equation}
with $m_\alpha$ the mass of the meson involved and $\Lambda_\alpha$ the so-called cut-off mass.'' Evidently, $\mathscr{F}_\alpha (m_\alpha^2) = 1$. It follows that, in order to obtain the equivalent values of the coupling 
constants $g_{\rho N N}$ and $g_{\omega N N}$ at $t=0$, one needs to multiply the quoted $g_\alpha^2/(4 \pi)$ values of Ref.~\cite{machleidt2001} by $(1 - m_\alpha^2 / \Lambda_\alpha^2)^2$, where (from Table I of that 
reference) $\Lambda_\rho=1.31$ GeV and $\Lambda_\omega=1.5$ GeV. One thus obtains the Bonn estimates found in Table \ref{tab:Constants} of this work. This point was obviously missed in Ref.~\cite{downum2006}; as a 
result, the value of $15.9$, quoted in the last column of their Table 1 as `the Bonn result', actually refers to the $\omega$-meson pole and, as such, it cannot to be compared with the values listed in their previous 
two columns. Following the definitions of Ref.~\cite{downum2006}, the correct Bonn estimate for $g_{\omega N N}$ (for Table 1 of Ref.~\cite{downum2006}) is $11.5$.

\begin{table}%[h!]
{\bf \caption{\label{tab:Constants}}}The reported values of the coupling constant $g_{\omega N N}$ and of the ratio $\kappa_\omega$, obtained in three standard $N N$ analyses. The table also contains the corresponding 
values of $g_{\rho N N}$ and $\kappa_\rho$ for comparison, as well as the ratio of the two vector couplings. The values of the coupling constants should be thought of as corresponding to $t=0$.
\vspace{0.2cm}
\begin{center}
\begin{tabular}{|l|c|c|c|c|c|}
\hline
Source & $g_{\omega N N}$ & $\kappa_\omega$ & $g_{\rho N N}$ & $\kappa_\rho$ & $g_{\omega N N}/g_{\rho N N}$\\
\hline
\hline
Nijmegen \cite{nagels1979} & $25.03$ & $0.655$ & $5.61$ & $6.602$ & $4.46$\\
Paris \cite{lacombe1980} & $24.30$ & $-0.12$ & $-$ & $-$ & $-$\\
CD-Bonn \cite{machleidt2001} & $23.07$ & $0$ & $4.22$ & $6.1$ & $5.47$\\
\hline
\end{tabular}
\end{center}
\vspace{0.5cm}
\end{table}

One notices that the reported values in Table \ref{tab:Constants} of the present work are not accompanied by any uncertainties; the frequent reply, whenever the relevant question arises, is that the dispersion-relation 
analyses are so fraught with theoretical constraints that the uncertainties in their output are a) difficult to estimate and/or b) make no sense. Be that as it may, using the three $g_{\omega N N}$ values of Table 
\ref{tab:Constants}, one may naively (average and unbiased estimation of the standard deviation of independent measurements without uncertainties) obtain:
\begin{equation} \label{eq:EQ06_1}
g_{\omega N N} = 24.14 \pm 0.99 \, \, \, .
\end{equation}
Lacking information on the uncertainty estimates in Refs.~\cite{machleidt2001,nagels1979,lacombe1980}, one should take a cautious attitude regarding the $g_{\omega N N}$ uncertainty in Eq.~(\ref{eq:EQ06_1}).

At this point, some insight into the interpretation of the physical constants $\kappa_\omega$ and $\kappa_\rho$ might be rewarding. The anomalous magnetic moment of the proton is put into the form $(1 + \kappa_p) \mu_N$, 
where $\mu_N$ is the nuclear magneton, a physical constant defined via the relation $\mu_N = e \hbar / (2 m_p) = 3.1524512550(15) \cdot 10^{-14}$ MeV T$^{-1}$ \cite{pdg2016}. The anomalous magnetic moment of the neutron is 
put into the form $\kappa_n \mu_N$. The current values of $\kappa_p$ and $\kappa_n$ are $1.7928473508(85)~\mu_N$ and $-1.91304273(45)~\mu_N$ \cite{pdg2016}, respectively. The isoscalar $\kappa_S$ and isovector $\kappa_V$ 
anomalous magnetic moments of the nucleon are defined as combinations of the physical constants $\kappa_p$ and $\kappa_n$: $\kappa_S = \kappa_p + \kappa_n \approx -0.12$ and $\kappa_V = \kappa_p - \kappa_n \approx 3.71$. 
In the VMD framework, $\kappa_S$ is identified with $\kappa_\omega$ (e.g., see the $\kappa_\omega$ value in the Paris potential \cite{lacombe1980} in Table \ref{tab:Constants}) and $\kappa_V$ with $\kappa_\rho$ 
\cite{sakurai1969}. Another remark is now due. It might appear reasonable to use $\kappa_V$ in the analyses of the $N N$ data, comprising measurements relating to all three $N N$ processes ($p p$, $p n$, and $n n$). On the 
contrary, the low-energy $\pi N$ measurements involve proton targets. Therefore, it might be more appropriate to use $\kappa_p$, rather than $\kappa_V$, in the $\rho N N$ vertex in the calculation of the contributions of 
the $\rho$-meson-exchange graph to the $\pi N$ scattering amplitude. Provided that this argument holds, a large part of the discrepancy, currently observed in comparisons between 
\begin{itemize}
\item the fitted $\kappa_\rho$ values in the PSAs of the low-energy $\pi N$ data using the ETH model and
\item the results obtained from standard dispersion-relation analyses,
\end{itemize}
could be understood.

\section{\label{sec:ScAmpl}Contribution of the $\rho^0 - \omega$ interference to the $\pi^\pm p$ ES amplitude}

\subsection{\label{sec:ScAmpl1}Literature on the $\rho^0 - \omega$ interference}

I will first discuss the literature (which I am aware of) on the $\rho^0 - \omega$ interference. In their 1987 paper, Coon and Barrett \cite{coon1987} determined the (EM) transition matrix element $\matrixel{\rho}{H}{\omega}$ 
and the potential $\Delta V^{\rho \omega}$ contributing to the violation of the isospin invariance in the $N N$ system. Using a new fit to the world data (at the time when the paper appeared) on the $e^+ e^- \to \pi^+ \pi^-$ 
reaction and a value for the branching ratio of the forbidden process $\omega \to \pi^+ \pi^-$ which is in conflict with the present-day knowledge, Coon and Barrett obtained $\matrixel{\rho}{H}{\omega} = (-4.52 \pm 0.60) \cdot 10^{-3}$ GeV$^2$, 
a result which was extensively used in subsequent works. Using the present-day values from Ref.~\cite{pdg2016}, one obtains $\matrixel{\rho}{H}{\omega} = (-3.39 \pm 0.13) \cdot 10^{-3}$ GeV$^2$; this value is almost identical 
to, and more precise than, the one obtained in an earlier paper by Coon, Scadron, and McNamee (see Ref.~[1] in Coon's and Barrett's paper). Utilising the (available at the time) data on the partial widths $\rho, \omega \to e^+ e^-$, 
and employing universality, Coon and Barrett obtained: $g_{\rho N N}^2/(4\pi) \approx 2.4$ (in reasonable agreement with the value given in Eq.~(\ref{eq:EQ05_1})) and $g_{\omega N N}^2/(4\pi) = 21.0 \pm 1.3$. One thus obtains 
$g_{\omega N N} = 16.24 \pm 0.50$; this value is problematic, as it corresponds to $t=m_\omega^2$ and should have been larger than the estimates of Table \ref{tab:Constants}. Coon and Barrett concluded that the $\rho^0 - \omega$ 
interference has the potential to account for most of the Okamoto-Nolen-Schiffer anomaly (i.e., for the binding-energy differences in mirror nuclei).

In their paper \cite{goldman1992}, Goldman, Henderson, and Thomas investigated the $t$ dependence of the $\rho^0 - \omega$ mixing amplitude, assuming that the interference is due to the $u$- and the $d$-quark mass difference; 
their quark model did not incorporate confinement. The authors found that the $\rho^0 - \omega$ mixing amplitude exhibits a strong $t$ dependence; this important finding inspired subsequent theoretical advancements. The same 
conclusion was reached by Piekarewicz and Williams \cite{piekarewicz1993}, who obtained an analytical expression for the $t$ dependence of $\matrixel{\rho}{H}{\omega}$, assuming that the $\rho^0 - \omega$ interference is 
generated by $N \bar{N}$ loops. In a paper which appeared in the same year, Krein, Thomas, and Williams \cite{krein1993} also obtained an analytical expression for the $t$ dependence of $\matrixel{\rho}{H}{\omega}$, on the 
basis of quark models incorporating confinement. In Ref.~\cite{cohen1995}, Cohen and Miller addressed a conceptual issue regarding the contributions of the $\rho^0 - \omega$ interference to the violation of the isospin 
invariance in the $N N$ system; they argued that the knowledge of the off-shell meson propagator is not sufficient to determine the $N N$ potential, and that a consistent approach would necessitate that the vertex functions 
be evaluated in the same theoretical framework which yield the propagator.

In a 1997 review article on the $\rho^0 - \omega$ interference, O'Connell, Pearce, Thomas, and Williams \cite{oconnell1997}, elaborating on their earlier ideas, established the `mixed-propagator approach' as the means to 
determine the contributions of the $\rho^0 - \omega$ interference to the scattering amplitudes. However, the approach was criticised by Coon and Scadron \cite{coon1998}, after they discovered that the violation of the 
charge symmetry, obtained in such an approach, is not consistent with the experimental results. The mixed-propagator approach was re-examined in a subsequent paper \cite{yan2002} and also found to be inconsistent, both 
in terms of the compatibility with effective Lagrangian models, as well as of the reproduction of the experimental data.

Finally, the propagator taking account of the $\rho^0 - \omega$ transition is also discussed in Refs.~\cite{azimov2003,azimov2003_2}. The interest in those studies was the extraction of the (on-shell) decay rates and of 
the decay-rate ratios of the $\rho^0$ and $\omega$ mesons into specific decay modes. As a result, the off-shell dependence of the functionals $G_{12}$ and $F_{12}$, pertaining to the transversal and to the longitudinal 
parts of the mixed propagator, is not addressed, the consequence being that these two articles, however didactical and reading with interest, cannot be of practical use in this work.

\subsection{\label{sec:ScAmpl2}Contributions to the $\pi^\pm p$ ES amplitude}

The contribution of the graph of Fig.~\ref{fig:IsospinBreakingRhoOmega} to the $T$-matrix element in the CM frame reads as:
\begin{align} \label{eq:EQ07}
\mathscr{T} &= -2 (-i)^4 g_{\pi \pi \rho} q^\prime_\mu \frac{g^{\mu\sigma}-k^\mu k^\sigma/m_\rho^2}{t-m_\rho^2} f(t) \frac{g_{\sigma\nu}-k_\sigma k_\nu/m_\omega^2}{t-m_\omega^2} \frac{g_{\omega N N}}{2} \bar{u}_f (p^\prime) \gamma^\nu u_i (p)\nonumber\\
&= - \frac{g_{\pi \pi \rho} g_{\omega N N}}{(t-m_\rho^2) (t-m_\omega^2)} f(t) \bar{u}_f (p^\prime) \fsl{q}^\prime u_i (p) \, \, \, ,
\end{align}
where the conventions of Ref.~\cite{matsinos2014} are used and $k$ denotes the $4$-momentum of the exchanged vector meson(s); $u (p)$ is the Dirac spinor associated with the plane-wave of a nucleon with $4$-momentum 
$p$, and the subscripts $f$ and $i$ refer to the final and the initial states, respectively. The function $f(t)$ denotes the $t$-dependent quantity $\matrixel{\rho}{H}{\omega}$. Therefore, the graph of Fig.~\ref{fig:IsospinBreakingRhoOmega} 
contributes only to the invariant amplitude $B$:
\begin{equation} \label{eq:EQ08}
B = - \frac{g_{\pi \pi \rho} g_{\omega N N}}{(t-m_\rho^2) (t-m_\omega^2)} f(t) \, \, \, .
\end{equation}

Normalised to the contribution (to the $B$ amplitude) of the vector coupling of the $\rho$-meson-exchange graph (see Section 3.2 of Ref.~\cite{matsinos2014}, in particular Eq.~(32) therein), the contribution of 
Eq.~(\ref{eq:EQ08}) becomes a measure of the violation of the isospin symmetry, imparted to the low-energy $\pi^\pm p$ ES amplitude as a result of the $\rho^0 - \omega$ interference of Fig.~\ref{fig:IsospinBreakingRhoOmega}:
\begin{equation} \label{eq:EQ08_1}
{\rm IV}_{\rho \omega} (t) = \frac{g_{\omega N N}}{g_{\rho N N}} \frac{f(t)}{m_\omega^2-t} \, \, \, .
\end{equation}

Results for ${\rm IV}_{\rho \omega} (t)$ will next be obtained after employing the two parameterisations of the $t$ dependence of $\matrixel{\rho}{H}{\omega}$ (i.e., the function $f(t)$ in Eq.~(\ref{eq:EQ08_1})) of 
Refs.~\cite{piekarewicz1993,krein1993}. To the best of my knowledge, these are the only papers where explicit, analytical expressions have appeared.

\subsubsection{\label{sec:ScAmplPW}Parameterisation of Piekarewicz and Williams \cite{piekarewicz1993}}

In first order (linear effects in the mass difference between the proton and the neutron), the $t$ dependence of $\matrixel{\rho}{H}{\omega}$ is contained in the expressions:
\begin{equation} \label{eq:EQ09}
f(t) = \frac{g_{\rho N N} g_{\omega N N}}{4 \pi^2} M \, \Delta M \, \left\{
\begin{array}{rl}
1 - \frac{1 + \xi^2 + \kappa_\rho}{\xi} \arctan \xi^{-1} & \text{, for} \, \, \, 0<t<4M^2\\
0 & \text{, for} \, \, \, t=0\\
1 - \frac{1 - \xi^2 + \kappa_\rho}{2 \xi} \, \ln \Bigl\lvert \frac{\xi-1}{\xi+1} \Bigr\rvert & \text{, for} \, \, \, t<0\\
\end{array} \right.
\end{equation}
where $M$ is the nucleon mass (average of the proton and neutron masses), $\Delta M$ represents the mass difference $m_n - m_p$, and
\begin{equation} \label{eq:EQ10}
\xi = \sqrt{\Bigl\lvert 1 - \frac{4 M^2}{t} \Bigr\rvert} \, \, \, .
\end{equation}
The factor of $4$ in the denominator on the rhs of Eq.~(\ref{eq:EQ09}) originates from the different conventions in the Lagrangian densities, followed in this work and in Ref.~\cite{piekarewicz1993}. For 
$\pi^\pm p$ ES, $t \leq 0$; therefore, the last two of Eqs.~(\ref{eq:EQ09}) apply.

\subsubsection{\label{sec:ScAmplKTW}Parameterisation of Krein, Thomas, and Williams \cite{krein1993}}

The $t$ dependence of $\matrixel{\rho}{H}{\omega}$, obtained in Ref.~\cite{krein1993} on the basis of $q \bar{q}$ loops, is contained in the expression:
\begin{equation} \label{eq:EQ12}
f(t) = \frac{3 g_{q \rho} g_{q \omega}} {16 \pi^2 \lambda^2} \exp \left( \frac{t}{2\mu^2} \right) \left( t + \frac{4}{\lambda} \right) \left( \frac{m_u^2}{m_d^2} - 1 \right) f_{0u}^2 \, \, \, ,
\end{equation}
where the various quantities are defined as follows \cite{krein1993}:
\begin{itemize}
\item $g_{q \rho}$ and $g_{q \omega}$ represent the coupling constants of the two light quarks ($u$ and $d$) to the $\rho$ and $\omega$ mesons respectively,
\item $m_u$ and $m_d$ stand for the `constituent' masses of the $u$ and $d$ quarks respectively,
\item $\lambda = 2 (\Lambda^{-2}+\mu^{-2})$, where the value of the cut-off constant $\Lambda$ is taken to be $1$ GeV, and
\item $\mu$ and $f_{0u}$ are parameters which may be obtained from the solution of the system of equations (copied from Ref.~\cite{krein1993}, Eqs.~(12) therein):
\begin{equation} \label{eq:EQ13}
f_{0u} = \frac{4 \pi^2 m_0^3}{3 m_u \mu^4}
\end{equation}
and
\begin{equation} \label{eq:EQ14}
F_\pi^2 = \frac{3 f_{0u} m_u^2 \mu^2}{4 \pi^2} \left[ 1 - \left( \alpha e^\alpha \int_\alpha^\infty \frac{dt \, e^{-t}}{t} \right) \right]_{\alpha=(m_u/\mu)^2} \, \, \, ,
\end{equation}
where the energy scale $m_0$ is related to the quark condensate ($\avg{q \bar{q}} \coloneqq -m_0^3$) and $F_\pi$ stands for the pion-decay constant.
\end{itemize}
Using the present-day values for the aforementioned quantities, namely $m_u=336$ MeV and $m_d=340$ MeV \cite{griffiths2008} (p.~135), $m_0 = 235$ MeV \cite{colangelo2001}, and $F_\pi = 130.2/\sqrt{2}$ MeV \cite{pdg2016}, 
one obtains $\mu \approx 553.89$ MeV and $f_{0u} \approx 5.40 \cdot 10^{-6}$ MeV$^{-2}$. Owing to the large difference of the $m_u$ and $m_d$ values between Refs.~\cite{krein1993} and \cite{griffiths2008} (the values of 
$450$ and $454$ MeV were used in Ref.~\cite{krein1993}), some attention is needed when determining the relevant contributions to the $\pi^\pm p$ ES amplitude.

Using the aforementioned current values of the physical constants leads to an unacceptable result for $\matrixel{\rho}{H}{\omega}$ at $t=m_\omega^2$, equal to about $-10.49 \cdot 10^{-3}$ GeV$^2$, i.e., nearly three times 
the result extracted on the basis of Eq.~(2) of Ref.~\cite{coon1987} (along with the present-day values of the physical constants). However, one notices that, in the parameterisation of Ref.~\cite{krein1993}, $f(t)$ is 
proportional to the product of the couplings $g_{q \rho} g_{q \omega}$. Although formulae for these two couplings (also containing the `constituent' quark masses $m_u$ and $m_d$) are given in Ref.~\cite{krein1993}, these 
expressions are approximate and, unavoidably, model-dependent.

In this work, an estimate for the product $g_{q \rho} g_{q \omega}$ will be obtained from the $\matrixel{\rho}{H}{\omega}$ value at $t=m_\omega^2$, extracted in Section \ref{sec:ScAmpl} after using Eq.~(2) of Ref.~\cite{coon1987}. 
The value of the product $g_{q \rho} g_{q \omega}$, thus obtained, is $6.47 \pm 0.26$ (as opposed to $20$, which was the recommendation of Ref.~\cite{krein1993}). This value of the product $g_{q \rho} g_{q \omega}$ makes 
the $\matrixel{\rho}{H}{\omega}$ estimates at $t=m_\omega^2$ of Refs.~\cite{coon1987,piekarewicz1993,krein1993} compatible\footnote{The estimate of Ref.~\cite{piekarewicz1993} for $\matrixel{\rho}{H}{\omega}$ at $t=m_\omega^2$ 
is $(-3.59 \pm 0.15) \cdot 10^{-3}$ GeV$^2$. This value has been obtained using $g_{\rho N N}$ of Eq.~(\ref{eq:EQ05_1}), $g_{\omega N N}$ of Eq.~(\ref{eq:EQ06_1}), and $\kappa_\rho=\kappa_V$ \cite{sakurai1969}. The use of 
the value of $6.1$ for $\kappa_\rho$ \cite{machleidt2001} yields a considerably lower result, about $-5.55 \cdot 10^{-3}$ GeV$^2$.}.

\subsection{\label{sec:Results}Implications of the $\rho^0 - \omega$ interference in the low-energy $\pi^\pm p$ ES}

\begin{figure}
\begin{center}
\includegraphics [width=15.5cm] {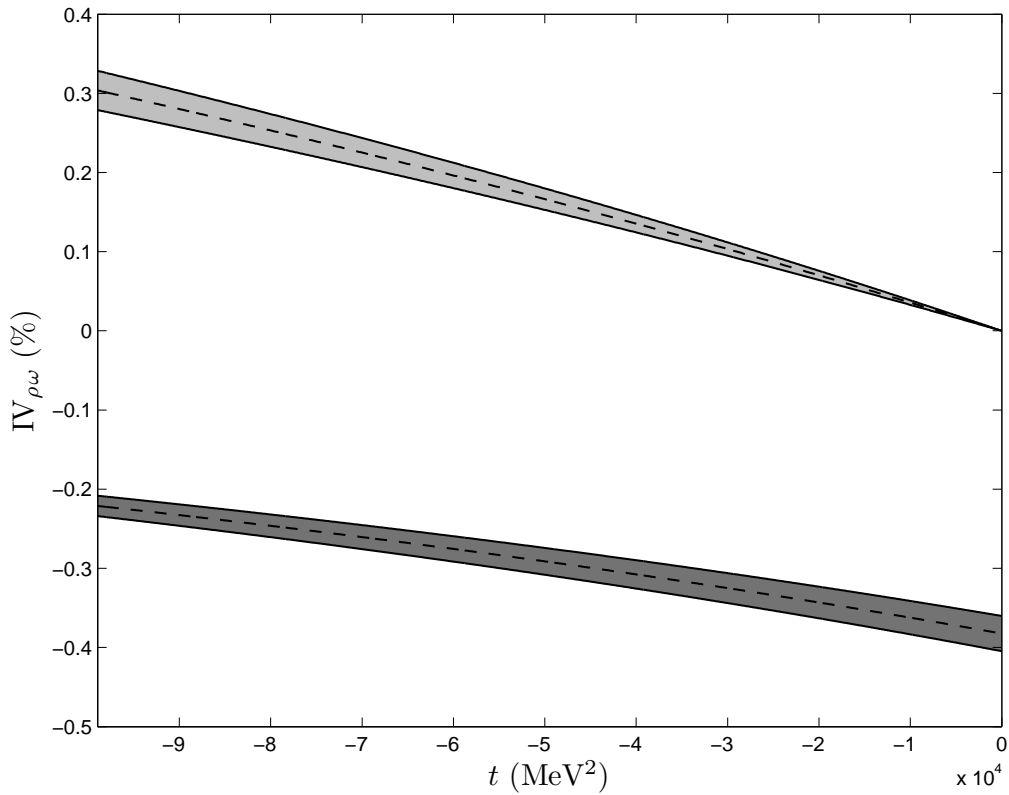}
\caption{\label{fig:IV}The fraction of the violation of the isospin invariance, induced on the invariant amplitude $B$ as a result of the $\rho^0 - \omega$ interference. The effect is shown as a function of Mandelstam's 
variable $t$, representing the square of the $4$-momentum transfer. The range of the $t$ values reflects the energy domain of the experimental data usually analysed with the ETH model. The bands represent the results 
obtained with the two parameterisations detailed in Sections \ref{sec:ScAmplPW} and \ref{sec:ScAmplKTW}; the upper band corresponds to the parameterisation of Piekarewicz and Williams \cite{piekarewicz1993}, the lower one 
to the parameterisation of Krein, Thomas, and Williams \cite{krein1993}. The shaded areas represent $1 \sigma$ uncertainties around the corresponding mean values (dashed curves); given that it is not straightforward to 
assign a more `realistic' uncertainty to the coupling constant $g_{\omega N N}$ (than the one quoted in Eq.~(\ref{eq:EQ06_1})), the $1 \sigma$ uncertainties in this figure might have been underestimated.}
\vspace{0.35cm}
\end{center}
\end{figure}

The results for ${\rm IV}_{\rho \omega} (t)$ of Eq.~(\ref{eq:EQ08_1}) for the parameterisations of Refs.~\cite{piekarewicz1993,krein1993} of the off-shell behaviour of $\matrixel{\rho}{H}{\omega}$ are shown in Fig.~\ref{fig:IV}. 
The range of the $t$ values in the figure reflects the energy domain of the measurements usually analysed with the ETH model ($T \leq 100$ MeV). A few remarks are due.
\begin{itemize}
\item The parameterisations of Refs.~\cite{piekarewicz1993,krein1993} lead to incompatible results for ${\rm IV}_{\rho \omega} (t)$, for $t \leq 0$. Evidently, they predict opposite signs for $f(t)$ in the low-energy region 
(this is also evident after inspecting Fig.~1 of Ref.~\cite{krein1993}).
\item Disregarding the sign difference, the measure of the violation of the isospin symmetry ${\rm IV}_{\rho \omega} (t)$ remains small in the low-energy region, at the few per-mille level.
\item $f(0) \neq 0$ in the parameterisation of Ref.~\cite{krein1993} (which is also obvious from Eq.~(\ref{eq:EQ12})).
\end{itemize}

It should be mentioned that, if the tensor-coupling contributions are also included\footnote{Unlike the vector coupling in the $\omega N N$ vertex, the tensor coupling also contributes to the invariant amplitude $A$; to 
obtain this contribution, one should simply multiply the contribution to the $B$ amplitude of Eq.~(\ref{eq:EQ08}) by $-\nu$, where $\nu$ is the standard Mandelstam variable representing the ratio $(s-u)/(4 m_p)$.}, the 
quantity ${\rm IV}_{\rho \omega} (t)$ of Eq.~(\ref{eq:EQ08_1}) should be multiplied by the ratio $(1+\kappa_\omega)/(1+\kappa_\rho)$, which would lead to even smaller ${\rm IV}_{\rho \omega} (t)$ (absolute) values 
than those of Fig.~\ref{fig:IV}.

\section{\label{sec:Conclusions}Discussion and conclusions}

One of the puzzling discrepancies in the analysis of the low-energy $\pi N$ data in this research programme \cite{matsinos2006,matsinos2014,ZRH17} involves the $\pi^- p$ elastic-scattering length $a^{cc}$ obtained via the 
extrapolation\footnote{This extrapolation is made on the basis of the ETH model (see Section 3.2.3 of Ref.~\cite{ZRH17} for details).} to the $\pi N$ threshold (zero kinetic energy of the incident pion) of the $\pi^- p$ 
elastic-scattering (ES) amplitude determined from the experimental data above threshold (and below pion laboratory kinetic energy of $100$ MeV). In a perfect world, this extrapolated result should come out compatible with 
the $a^{cc}$ values extracted from the strong-interaction energy-level shift $\epsilon_{1s}$, determined to high accuracy in two experiments on pionic hydrogen, conducted at the Paul Scherrer Institut \cite{schroeder2001,hennebach2014}.

Corrected for electromagnetic effects according to Ref.~\cite{oades2007}, the $a^{cc}$ value of Ref.~\cite{schroeder2001} is equal to $0.08576(80)~m_c^{-1}$, the one of Ref.~\cite{hennebach2014} is $0.08549(59)~m_c^{-1}$, 
where the statistical and the systematic uncertainties in these experiments (properly reported in the original papers) have been linearly combined herein for convenience\footnote{In the domain of $\pi N$ Physics, the 
scattering lengths are routinely expressed in $m_c^{-1}$, rather than in fm.}. Therefore, the two `experimentally-obtained' $a^{cc}$ values are perfectly compatible with one another.

On the contrary, the extrapolated $a^{cc}$ value is $0.0799(11)~m_c^{-1}$ \cite{ZRH17}. As a result, the difference between the two scattering-length values is $0.0059(14)~m_c^{-1}$ in the case of the original pionic-hydrogen 
experiment \cite{schroeder2001}; concerning the follow-up experiment \cite{hennebach2014}, the difference amounts to $0.0056(13)~m_c^{-1}$. One might correct the extrapolated $a^{cc}$ value for energy-dependent effects, 
present in the optimal scale factors of the fits to the low-energy ES data, in which case the aforementioned differences reduce to $0.0047(16)~m_c^{-1}$ and $0.0045(15)~m_c^{-1}$ for Refs.~\cite{schroeder2001} and 
\cite{hennebach2014}, respectively; at this time, there appears to be no way, by means of which the residual difference could be eliminated (or further reduced).

The closest to the experimental results the model predictions have ever reached was in Ref.~\cite{matsinos1997}, where robust statistics was employed in the optimisation and the floating of the experimental data sets was 
not allowed; the $a^{cc}$ value of Ref.~\cite{matsinos1997} (p.~3022) was: $0.0826(20)~m_c^{-1}$. The main conclusion of the present work is that, if the mixing amplitude $\matrixel{\rho}{H}{\omega}$ is as small in the 
vicinity of the $\pi N$ threshold as Refs.~\cite{piekarewicz1993,krein1993} predict it to be, the residual $3 \sigma$ difference in the $a^{cc}$ values can hardly be attributed to the violation of the isospin symmetry due 
to the $\rho^0 - \omega$ interference.

The second motivation for this work was the provision of justification or refutation of the strategy followed in Refs.~\cite{gibbs1995,matsinos1997} in claiming that reliable information may be extracted from the $\pi N$ 
system in a combined analysis of the two ES reactions. The inevitable consequence of Fig.~\ref{fig:IV} is that the strategy followed in these analyses is justified. The magnitude of the violation of the isospin symmetry, 
originating from the $\rho^0 - \omega$ interference according to Refs.~\cite{piekarewicz1993,krein1993}, is well below the experimental uncertainties of the measurements in the low-energy region; only the normalisation 
uncertainties of the available data sets are usually around $3 \%$.

The results for $f(t)$ of Eqs.~(\ref{eq:EQ09},\ref{eq:EQ12}), obtained on the basis of Refs.~\cite{piekarewicz1993,krein1993}, are opposite in sign in the low-energy region; these two references agree only on the overall 
significance of the isospin-breaking effects imparted to the $\pi^\pm p$ ES amplitudes as a result of the $\rho^0 - \omega$ interference. Of course, this disagreement is not an artefact of the rescaling of the product 
$g_{q \rho} g_{q \omega}$ of Ref.~\cite{krein1993}, discussed in Section \ref{sec:ScAmplKTW}; the parameterisation of Ref.~\cite{piekarewicz1993} is compatible with $f(0)=0$ (i.e., with the expectation of Ref.~\cite{oconnell1994}), 
whereas that of Ref.~\cite{krein1993} results in $f(0)<0$. Last but not least, the inclusion of form factors in the parameterisations of Refs.~\cite{piekarewicz1993,krein1993} in the low-energy region does not alter the 
conclusions of this work.

\begin{ack}
I am indebted to R.~Machleidt for clarifying a question regarding Ref.~\cite{machleidt2001}. Also acknowledged is an interesting exchange of e-mail with G.A.~Miller.

The Feynman graphs of this paper were drawn with the software package JaxoDraw \cite{binosi2004}, available from jaxodraw.sourceforge.net. Figure \ref{fig:IV} has been created with MATLAB \textregistered~(The MathWorks, Inc.).
\end{ack}

\end{document}